# On Systemic Destruction of Human Locomotor System

Zoja Medjanik[1], Lyudmila Popova[1], Sergiy Popov[2]*

[1]Medik BT, Veszprem, 8200, Hungary.

[2]Control Systems Research Laboratory, Kharkiv National University of Radio Electronics, Kharkiv, 61166, Ukraine.

*Correspondence to:
Dr. Sergiy Popov.
Ph.D., Dr.Sc., IEEE Senior Member
Chief Researcher
Control Systems Research Laboratory
Kharkiv National University of Radio Electronics
14, Lenin av., r. 517
Kharkiv, 61166, Ukraine

E-mail: Serge.Popov@gmx.net

## Abstract

Locomotor system disorders affect a vast majority of people at some time in their life bringing pain, functional limitations, social and economic implications. Modern medicine cannot offer prevention and effective treatment for most chronic musculoskeletal conditions, because their etiology and pathogenesis are unknown. This is due to the lack of systemic understanding of the locomotor system functioning in both healthy and unhealthy states. Here we apply systems sciences to analyze the human locomotor system and develop a general theory that reveals the systemic destructive process in the locomotor system, linking together all its disorders. The systemic destruction involves adaptation and self-organization processes in the locomotor system, whose side effects introduce a positive feedback loop with nervous and vascular disturbances. Most chronic musculoskeletal conditions are just manifestations and consequences of this process. On the basis of our theoretical findings, we developed the world's first technology that effectively counteracts the systemic destruction and improves the locomotor system state at any age, also preventing problems in nervous and cardiovascular systems.

## Introduction

Analysis of a broad spectrum of literature and 30 years of our research and practical work with patients from many countries have shown that orthopedics and related branches of medicine are currently in a severe theoretical and practical crisis. It is manifested by the lack of theory of etiology and pathogenesis of prevalent chronic musculoskeletal conditions, which already top the list of major causes of disability. These include:

– degenerative conditions in the spine and joints: osteoarthritis, degenerative disk disease, etc. [1,2];

– skeletal deformities: scoliosis, kyphosis, etc. [3];

– many types of muscle pathology: muscular dystrophy, myofascial trigger points, fibromyalgia, etc. [4,5,6];

– back and neck pain, and other musculoskeletal pain with an indeterminate etiology [7,8].

The lack of theoretical foundation results in a practical inability of modern medicine to effectively treat these disorders. Here are just a few facts. Musculoskeletal conditions are the most common cause of severe long-term pain and physical disability, and they affect hundreds of millions of people around the world. They cause more functional limitations in the adult population in most welfare states than any other group of disorders. They are a major cause of years lived with disability in all continents and economies [9]. Almost everyone has structural evidence of osteoarthritis on radiographs or magnetic resonance imaging in at least one joint by 70 years of age [10]. Low back pain has reached epidemic proportions, being reported by about 80% of people at some time in their life [11]. Americans spend at least $50 billion each year on low back pain, the most common cause of job-related disability and a leading contributor to missed work [8,12]. At any one time, 30% of American adults are affected by joint pain, swelling, or limitation of movement [9]. With annual increase of the number of patients needing care, the situation is even more aggravated by the shortage of surgeons and resources [13]. Under these conditions the question of finding causes and elimination of visual body imperfections (which are not medical pathologies yet) is not even raised, in spite of its high importance for everyone from an aesthetic perspective.

At the turn of the millennium the situation became so threatening that a group of health professionals supported by the World Health Organization, the World Bank and the United Nations decided to start a global campaign, declaring the first 10 years of the $21^{st}$ century the "Bone and Joint Decade" (BJD) [14]. The initiative has been endorsed by 48 national governments and more than 750 national and international patient and scientific organizations, and related journals. The BJD aims to help keep people moving by promoting the prevention and treatment of musculoskeletal disorders. To achieve this, several goals have been determined, among which is to advance understanding of musculoskeletal disorders through research on improving prevention and treatment [15]. A research project was implemented to measure the health impact and economic burden of musculoskeletal conditions. The analysis of the US statistical data has shown that the cost of musculoskeletal conditions would appear to be growing with the aging of the population and the increased utilization of new medical technologies. In 1995 it was approaching 3% of GDP (US$ 214.9 billion) that is equivalent to a permanent severe recession [11]. By 2004 this number increased by 2.5 times and reached 7.7% of GDP (US$ 849 billion) [16]. As the prevalence of many musculoskeletal disorders increases with age, the global situation is likely to worsen in the future.

The Bone and Joint Decade which reached completion in September 2010 was deemed to be a great success [17] and of such vital importance that is has been extended for another decade [18]. However, little was achieved towards the aforementioned goal of advancement of understanding of musculoskeletal disorders through research on improving prevention and treatment. The advancements are limited and local from the diagnoses perspective, and no uniform approach to the assessment and management of these disorders has been discovered yet [19]. We strongly believe the main reason for this to be the lack of systems approach to the analysis of musculoskeletal disorders. Consequently, for each type of pathology (there are more than 150 different musculoskeletal disorders [15]) its own causes are sought, independent etiology and pathogenesis theories, and specific treatment methods are developed. In isolation from common, systemic principles of functioning of the musculoskeletal system, these theories prove inconsistent, and the corresponding treatment is ineffective.

Therefore it is vital to take a broader look at the aforementioned problems particularly and the musculoskeletal system functioning in general using systems sciences (including general systems theory, control theory, optimization theory), which would enable development of a general theory describing processes in the musculoskeletal system. It is this vision of the musculoskeletal system's pathology problem that can make a qualitative leap and open up entirely new perspectives for a successful treatment. It should be noted that some attempts to apply systems approach to analyze musculoskeletal system's functioning were made occasionally since the 1960's [20,21], however to date they did not lead to significant advancement in understanding of its pathology, less to practical solution of the problems listed in the beginning of the paper.

## Theoretical background

In terms of systems sciences, the human body is a very complex biological system consisting of a number of subsystems: nervous, cardiovascular, digestive, etc. Each of them in turn can be represented as a set of subsystems of different levels: individual anatomic units, tissues, cells, and all of them again will be very complex biological systems. Despite the enormous complexity of the human body and the processes inside of it, it is able to survive in a broad range of environmental conditions. This is possible only due to the purposeful control by its control system – the nervous system, whose center is the brain. This control is implemented on a subconscious level, independently of human consciousness and will.

The human body constantly experiences the action of mechanical forces: gravitation, atmospheric pressure, supporting surface reaction, friction, work load, etc. To counteract them and to move itself and other objects in space, the human body contains a complex biomechanical system – the Locomotor System (LS). It is a living mechanical device (biomachine) that experiences, distributes, and itself generates mechanical forces. The LS consists of the Musculoskeletal System (MS) – the "plant" (in terms of the control theory), and parts of the nervous system responsible for the control of the musculoskeletal system – the Locomotor Control System (LCS). In most cases, there is no point in considering the MS independently, without its control system, because it is the LCS that is responsible for maintaining a certain position and for movement of the body in space, which are the LS system functions. Without continuously acting control signals, muscles will completely relax, and the MS will seek to take a position with a minimal potential energy of its elements, i.e. it will simply fall. For this reason, in our research we consider the locomotor system as a whole (implying that the LCS is intact), and not the musculoskeletal system alone or its separate elements, as it is done in most other studies.

The LS is a subsystem of the human body, and the LCS is a subsystem of the whole Body Control System (BCS). The concept of control necessarily involves some actions aimed at achieving certain goals, either explicitly or implicitly defined [22]. Thus, consideration of any aspect of the LS requires its representation as a purposefully controlled system. Modern science studies various aspects of the LS structure and functioning: mathematical and mechanical models of the MS are built [23,24], analogies between biological and technical control systems are drawn [25], locomotion control is considered [26,27], stability and control of separate MS segments are studied [28,29,30], etc. However there is an evident lack of a general, systems approach to studying the LS in a healthy, unhealthy, and diseased states, and most important – a general mechanism of transition between these states is completely overlooked. This paper reveals and describes this universal mechanism that covers all types of LS pathologies. To achieve this, we analyze LS functioning and pathology in a general form, abstracting from its structural details and specific diagnoses. To this end, we do not pay any particular attention to the spine, which is an ordinary LS unit in this context.

## Basic concepts and definitions

To proceed, we need to introduce several new concepts.

The LS anatomic state is the constellation of states of all LS elements (anatomic units):
− their spatial position,
− composition and properties of their tissues,
− their shape.

The normal LS anatomic state (the reference LS state) is the state when every LS element (anatomic unit):
− takes the anatomically normal position in the LS spatial structure,
− has the anatomically normal composition and properties of its tissues,
− has the anatomically normal shape.

When there is a persistent deviation in any of these properties in at least one LS element (or any number of them in any combination) the whole LS is said to be in a disturbed state. We call the LS in this disturbed state the pathologically altered or destructed LS, and the process of emergence of these deviations (destructive changes) is called the LS destruction. The term "persistent" means that the organism itself cannot restore the normal state of the element. A persistent deviation in the element's spatial position is called a displacement.

Loads acting on the LS may be either static or dynamic, and may vary in a broad range, e.g. carrying heavy objects, running, hanging by hands, falling, etc. We consider only static LS behavior, because the LS anatomic state can be unambiguously assessed only in a static mode. We use the primary position for certainty: an individual stands upright on a horizontal surface on straightened legs without additional support and without conscious muscle tension, the load is distributed evenly on both feet, the head is in the straight position, the arms are hung freely, the centers of mass of the head and the whole body, and the lateral axes of the large joints of the limbs lie in one plane.

We call static loads on the elements of the reference LS in the primary position the anatomic loads. Those are the loads predetermined by the LS design and mechanical properties of its elements in the normal LS anatomic state. Consequently, nonanatomic loads are the loads on the LS elements that differ in any way (in absolute value and/or direction) from the anatomic loads, i.e. nonanatomic loads are not inherent by design to the LS in the reference state. Nonanatomic loads are persistent, if the organism itself cannot eliminate them (because of their organic origin, e.g. trauma or degenerative conditions) or if they are sustained by the LCS on a subconscious level (muscle hypo- and hypertonicity).

Since most of the LS elements are mobile, their spatial position is determined by the forces acting on them. Therefore concepts of the normal LS anatomic state and of the anatomic loads are closely related: if the LS is in the normal anatomic state, then all of its elements experience anatomic loads, and vice versa, if all the LS elements experience anatomic loads, then the LS anatomic state is normal. If there are nonanatomic loads (regardless of their origin) in the LS, their balancing is achieved in another (not anatomically normal) position of bones and other LS elements, and muscle tone is not normal. Another important concept is a permissible load. If the load was permissible, the LS can by itself return to the reference state after the load is removed, i.e. a permissible load does not cause persistent deviations in spatial position, shape and properties of LS elements. When the LS experiences any nonanatomic load, it comes out of the reference balanced position. To maintain stability, the LCS compensates that load by changing the muscle tone without the individual's will, hence the LS takes a new stable position in space – this is called adaptation. If the nonanatomic load lied in the permissible range, when it is removed, the LCS returns the LS to the original reference position.

Every individual's actual LS anatomic state at any stage of life always more or less deviates from the reference (normal LS anatomic state). Moreover, the LS destruction progresses with age, deviations from the normal anatomic state increase due to the utilization of the LS as a mechanical device, systemic disorders in the whole body and its ageing, which involves pathological changes in the whole organism, if the pathology is termed as a deviation from the norm. Thus, the LS destruction is an integral part of the ageing process. The more the actual LS anatomic state deviates from the reference, i.e. the more the LS destruction advances, the more abnormal are the spatial configuration, composition and properties of the body tissues, the higher the probability of disorders onset, the higher the LS functional impairment, because the LS functioning directly depends on its structure and control quality. Hence, the problem – the difference between the desired and the actual states in terms of systems analysis [31] – is that the actual LS anatomic state always deviates from the normal LS anatomic state, because of the destructive changes in the LS. Therefore, it is desirable for every individual to have as little deviations from the reference LS state as possible.

**Locomotor system control objectives, stability, muscle tone**

To study the LS functioning as a purposefully controlled system, it is important to understand that the LCS is continuously solving a multi-objective optimization problem. In most cases its objectives are conflicting: improvement of one objective leads to suffering of the others. In such a case, when it is impossible to improve all objectives simultaneously, a ranking mechanism is employed – some objectives are prioritized over the others. We believe that under normal conditions the LCS objectives are prioritized in the following order. The principal objective is minimizing energy costs, because the energy loss directly influences the organism viability. That is why every subsystem's control system in the organism regardless of its local objectives obeys the general energy costs minimization objective. The second important objective of the LCS is spatial stability, since without stability the LS is unable to accomplish any useful task. Minimizing the negative impact on LS elements is in the third place, pain is the signal of an intolerable level of this objective. And the success of the current locomotor task execution takes the last place.

It should be noted that at certain moments this ranking may be consciously or unconsciously altered to achieve a higher-level objective, unknown to the LCS, or due to changes in the operating conditions of the body. The significance of energy costs minimization objective depends on the currently available energy reserve. If this reserve is high enough, the BCS may allocate a significant amount of energy to optimize other objectives. If the body is low on energy, then the BCS switches to the austerity mode and may even disable certain body functions completely [32]. The minimization of energy costs and the stability objectives are conflicting by their nature, since the latter requires an increase in muscle tension, which leads to a deterioration in the former. In the optimization theory in such cases, one of the objectives is transformed into a constraint; in our case, achievement of a certain level of stability (sufficient for normal body functioning) is a constraint, while energy costs are minimized. The optimum for this objective is achieved, when the LS is in the normal anatomic state.

With these principles in mind, let us consider the LCS system goal – ensuring the LS spatial stability, which is a necessary condition for its efficient functioning. Stability is defined as a system's property to change its state little under the influence of perturbations; for a more rigorous treatment of the stability concept refer to the fundamental works of A.M. Liapunov [33]. From a mechanical perspective, the MS is a set of interconnected elements of several types: bones, cartilages, tendons, ligaments, fasciae, and muscles. Muscles are the only active LS elements controllable in an "online mode", all the other anatomic units are passive. Considering the LS in the primary position without active elements (muscles), we can see that this structure is unstable, because its center of mass is located high (about 1 meter) above the supporting surface, whose area is only several square centimeters (parts of the feet contacting the support), and all main LS segments are mobile with many degrees of freedom. In a significantly simplified form this structure can be represented as a multisegment inverted pendulum [34]. It can be balanced, however even a slight deviation (perturbation) from the equilibrium position will lead to its fall (significant change of state). Hence, to achieve stability, it is necessary to limit mobility of LS elements. This is done by active, controllable elements – muscles under the control of the LCS.

Thus, the LS spatial stability is achieved only due to its control system [35]. The LCS continuously monitors the LS spatial position and controls it using active elements – muscles. The complexity of this process is hard to imagine. In a fraction of a second the LCS processes information from thousands of position sensors in different parts of the body, from tactile sensors in the skin that inform about the external pressure, processes visual and vestibular data, makes a decision and generates commands individually for each of several hundred muscles. All this takes place in real time under continuously changing environmental conditions. To be stable is as simple and usual for humans, as breathing. In fact, behind this apparent simplicity lies a very complex unconscious control algorithm. So far, medicine did not pay due attention to searching for the reasons of LS pathologies by studying the functioning of this algorithm.

In the reference LS, in primary position at rest loads are distributed over its bearing elements: bones, cartilages, ligaments, and fasciae, which do not consume energy to carry loads. The role of muscles then reduces to maintaining balance, i.e. instantly returning the LS to the equilibrium position, if it deviates from it due to perturbations. For the equilibrium position to be stable and insensitive (robust) to numerous small perturbations (breathing, heartbeat, etc.), muscles must be in a state of small tension (muscle tone). The LCS maintains this tension subconsciously, irrespective of the individual's will, i.e. fixes the LS in the equilibrium position. This mode of operation is optimal from an energetic perspective, since the load is distributed over the elements, which do not consume energy to carry it. Hence, in the reference LS state, energy costs of maintaining a stable position are minimal: this energy is needed to continuously maintain the normal muscle tone. The normal muscle tone is the muscle tension, which provides static stability of the LS with the normal anatomic state in the reference position without additional loads. It must be well understood that muscle tone is the same muscle tension as any other and it also consumes energy. The difference is that, unlike purposeful tension, the LCS maintains muscle tone without the individual's will.

## Main result

Now, using the above theoretical foundation, consider the mechanism of changes in the LS, by means of which it leaves its reference state. As the normal LS anatomic state is defined as the state with reference characteristics over three types of parameters, there are three types of deviations for any LS element and for the whole LS:
− a persistent deviation from the reference LS spatial structure that corresponds to a persistent deviation from the anatomically normal position (displacement) of at least one LS element,
− a persistent deviation from the anatomically normal composition and properties of LS elements tissues,
− a persistent deviation from the anatomically normal LS elements shape.

By definition, if there is even a single persistent deviation, not only that LS element, but the whole LS is deemed destructed, i.e. pathologically changed. At the same time, in this destructed LS anatomic state some LS elements may have no deviations.

There are many causes independent of the LS that lead to persistent deviations from the normal LS anatomic state. Altogether we call them Initiating and Intensifying Factors (IIFs) of the systemic destructive process. Here is a list of main IIFs (we do not claim it to be complete, the more that the specifics of these factors is not principal for our study):
− congenital LS anomalies and LS tissues diseases;
− various injuries (including birth injuries) and microtrauma;
− surgical and other nonsystemic mechanical interventions;
− violations of the optimal motor mode: prolonged static loads (also due to excess weight), prolonged dynamic overloading, lack of physical activity;
− disorders in other body systems: cardiovascular, nervous, etc., diseases of internal organs;
− tumors and any other diseases of LS tissues, infectious diseases and inflammatory processes in the LS and the whole body;
− pain in various locations, causing the muscle response and spasm;
− endogenous and exogenous damaging agents (chemicals, electromagnetic radiation, etc.);
− metabolic disorders, intoxication;
− handedness, i.e. a predominant use of one hand (right for right-handed individuals, left for left-handed individuals);
− the natural ageing process.

Some of these factors (e.g. an injury) initiate persistent nonanatomic loads immediately following the external impact and instantly disturb the LS from an equilibrium position. The others (e.g. a chronic LS tissue inflammation) may act incrementally, causing the anatomic loads on LS elements to change gradually into persistent nonanatomic loads, disturbing the body load balance. Prolonged and daily recurring static and dynamic loads may cause persistent hypertonicity in certain muscles with simultaneous hypotonicity in their antagonists. i.e. persistent nonanatomic loads, which in turn cause displacements of bones and other LS elements. Thus, analysis of all possible factors and types of deviations reveals that any persistent deviation from the normal LS anatomic state initiates persistent nonanatomic loads in the whole LS (the whole system load balance changes due to its systemic mechanical properties), which influences the LS stability.

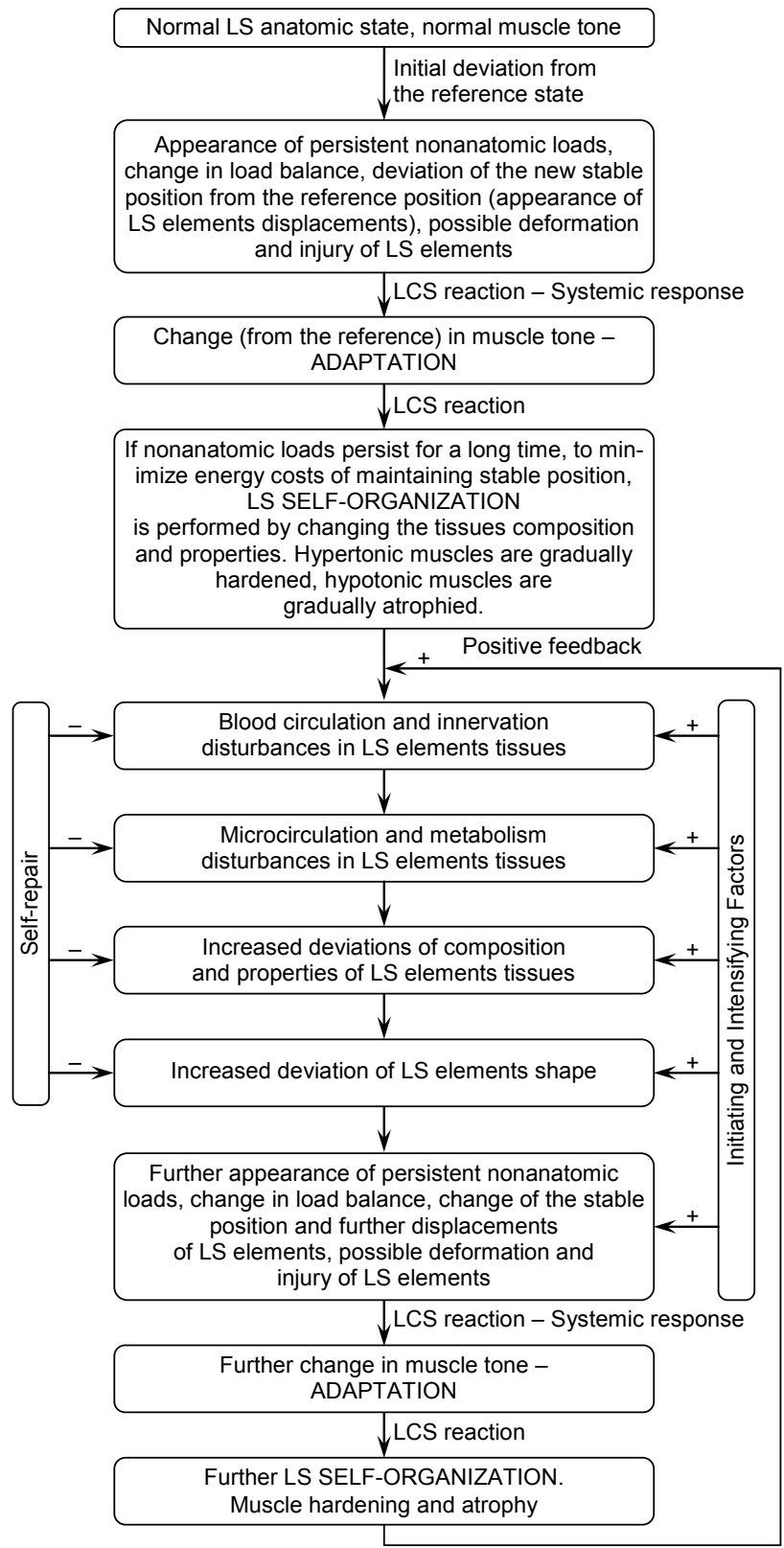

Figure 1. A simplified diagram of the systemic destructive process in the locomotor system

The process of the LS anatomic state change (when it leaves the reference state and all further changes during the individual's life due to recurring deviations for different reasons) can be conveniently represented in a simplified diagram (Fig. 1). Please note again that we consider

LS functioning and pathology in a general form. This abstraction from specific conditions and diagnoses made it possible to fully employ the systems approach and reveal the true nature of the variety of destruction processes in the locomotor system, which unfold in space and time during the whole individual's life.

The first deviation of the LS anatomic state from the reference state, and all subsequent deviations (both nonsystemic ones due to IIFs, and systemic ones caused by the systemic destruction) lead to changes in the load balance in the whole LS, which may cause the loss of stability. To maintain stability, the LCS performs adaptation: it changes the muscle tone in the whole LS (because every LS element influences the stability of the whole system), consequently, bones connected to these muscles are displaced from their normal positions within the range of joints free motion. As far as this process affects the whole system, we call it the systemic response.

A changed muscle tone (in active individuals it is predominantly hypertonicity) implies increased energy costs. To minimize them, the LCS eventually fixes the changed muscle tone with the organic changes (degeneration) of the muscle tissue – the process of self-organization (structural adaptation). Hypertonic muscles gradually harden, taut bands are formed, which generate the necessary permanent force not by tension, but due to elastic stretching, hence without energy consumption. Organic degeneration of muscle tissue cannot be reversed by the human organism itself. It is a destructive change in the LS, as it (by definition) disturbs anatomically normal composition and properties of muscle tissues. A decreased muscle tone – hypotonicity, which is observed in most muscles of bed-patients or under severe hypodynamic conditions (e.g. due to a prolonged period in weightlessness), is also fixed with the organic degeneration: the LCS subjects the unloaded muscles to atrophy. In this case, energy costs are reduced by decreasing the energy supply to the muscles, which are no longer needed in their fully fledged state. Atrophy is also a destructive change of muscle tissues. Muscle hardening and atrophy are the main escalating factors of the LS systemic destruction.

We would like to stress an extremely important point, which is new to medicine: these organic changes in muscles take place not because of some reasons, but are purposefully initiated by the LCS to reduce continuous energy consumption of the human body. By self-organization, the body modifies itself to survive more efficiently. The fact that muscle atrophy can take place not only because of their underutilization, but can be initiated by the LCS due to lack of support afferentation was discovered in space medicine back in the 1970's [36], however the connection with the LCS control objectives (particularly, energy costs minimization) had not been revealed. The opposite process – the purposeful muscle hardening that substitutes chronic hypertonicity – has never been described previously. Its discovery is the result of our many years' practical work with thousands of patients with various LS pathologies. It is important to understand that both these phenomena (muscle hardening and atrophy) are manifestations of the same general LS self-organization process.

Organic changes in muscles and nonanatomic loads on bones and other LS elements disturb nervous and vascular conductivities in both these and adjacent elements. This consequently disturbs microcirculation and metabolism in LS elements tissues, which eventually leads to increase of destructive changes in the already destructed LS elements and appearance of such changes in adjacent elements that were normal prior to this moment. This way, new (systemic, i.e. caused by the processes inside the system) deviations appear in the LS, which again change the LS load balance, initiate a new stage of adaptation and self-organization, so introducing a positive feedback loop into this process. Thus, the adaptation and self-organization processes caused by the very first initiating nonsystemic deviation exhibit a positive feedback with system-wide disturbances in nervous and vascular conductivities. This "vicious cycle" leads to more and more new destructive changes in the LS, which accumulate and reinforce themselves (much like chemical and nuclear chain reactions). Eventually, this process takes a systemic nature and progresses in time and space even without external (independent of the process itself) influence. During the whole individual's life the LS anatomic state increasingly deviates from the normal LS anatomic state, pathology gradually spreads and generalizes on the whole LS. We call this discovery the LS Systemic Destruction (SD).

The SD disturbs the LS structural integrity. Many non-bearing elements (muscles) become bearing ones that limits the individual's motor activity, while bearing elements (bones, ligaments, etc.) properties partially degrade that weakens the LS support function. At advanced SD stages, LS elements, which are not sufficiently nourished through the organically degenerated muscles, are severely affected: bones (osteoporosis), tendons (tendinitis), cartilages (osteoarthritis); joints range of motion is restricted accordingly. This restriction is caused neither by joint locks due to muscle spasm (muscles cannot have chronic spasm as this contradicts the energy costs minimization objective, spasmed muscles eventually exhibit organic degeneration), nor by disturbance of the central nervous regulation (muscle imbalance), nor by meniscoid entrapment or joint subluxation; it is a consequence of organic pathology in the whole LS. That is why such conditions cannot be eliminated by massage, acupuncture, CNS retraining with exercise therapy or kinesiotherapy, instant manipulations, etc., which explains ineffectiveness of these methods in treating chronic LS pathologies [7].

The LS systemic destructive process is opposed by the natural self-repair process in the body that consists in adaptation and self-organization of its other subsystems, which are negatively affected by the SD (nervous and cardiovascular systems in the first place). The self-repair decelerates the SD, however due to limitation in the body resources it cannot stop the SD completely. On the other hand, IIFs emerge and keep acting during the whole individual's life that introduces additional nonsystemic deviations and accelerates the SD.

Thus, any pathology (deviation from the reference state regardless of its specific nature) in any LS element disturbs functioning of the LS as of a mechanical device that affects its stability in the first place. The LCS continuously performs its main function – maintains body spatial stability, it reacts to emerging pathologies by adaptation (systemic response). If the pathology is not eliminated quickly, the LCS modifies the LS structure in the process of self-organization obeying the predominant energy costs minimization objective. This algorithm of reaction to pathology is developed by millions of years of evolution of the man and his ancestors. Neither the LCS, nor control systems of the other body subsystems, nor their coordinator (the higher level control system – the BCS) never anticipate any external help, e.g. medical assistance, and are even unaware of its existence at all, since their control of the human body is performed subconsciously, in an automatic mode.

In the course of systemic destruction, destructive organic changes accumulate in the LS. They gradually make the LS go out of the healthy state that is manifested by visual body imperfections, discomfort, etc. After a certain verge is crossed, clinical symptoms appear: pain, inflammation, cartilage tissue degeneration, etc. At this stage such a condition is categorized as a disorder and only now it enters the medical field of view. Body mobility and ability to perform heavy physical work are progressively degraded that may eventually lead to disability, i.e. LS functional failure.

LS deformities develop according to the same scenario: from various body imperfections inherent to almost every human to severe curvatures, which begin in the childhood and are manifestations of early SD generalization (kyphosis, scoliosis of higher degrees). In this respect, the theory of the LS systemic destruction may be considered as the universal LS deformation theory, which modern orthopedics lacks so much. Not seeing the specific causes of pathology and not knowing about the SD process producing it, physicians consider these disorders as independent, multifactorial, with an indeterminate etiology. The theory of the LS systemic destruction presents new and crucial for the clinical medicine concept of the systemic degeneration (profound organic change) of the LS structure, which develops from the first years of the individual's life and usually precedes spinal disorders. Moreover, most spinal disorders (except injury cases) are consequences of the SD.

**Discussion**

As the result of many years of practical research we conclude that, despite the apparent difference in the prevalent chronic musculoskeletal conditions, all of them are interconnected by the single systemic destructive process in the locomotor system and they are just its evident manifestations. The SD is the major cause of intervertebral disk herniation, spinal instability, joints functional limitations, musculoskeletal pain (including back pain and myofascial trigger points), scoliosis, kyphosis, simply anatomically incorrect (distorted) body contours, disproportions of its parts, and many consequent neurological and vascular disorders that are manifested by headaches, abnormal blood pressure and heart rate, tissue ischemia, strokes, heart attacks, impaired function of internal organs, mental disorders, chronic fatigue syndrome, etc. The SD can cause muscle atrophy, osteoporosis, premature skin ageing on the face and body, wrinkles formation. An individual with the SD in a generalized stage constantly feels the "burden" of his/her body and chronic discomfort that worsen with physical activity and in the morning. The SD phenomenon also manifests itself by a persistent, chronic clinical course of the disorder, wherever it is located, by regularity of exacerbations and its practical incurability by all known methods. And no matter which specific diagnosis a patient has, which nerve roots are entrapped or which joints have degenerative changes – the pathology pursues the patient and aggravates during the whole his/her life.

Because there are a lot of initiating and intensifying factors of the SD and some of them appear in life of every human, it is impossible to avoid the SD. The only thing an individual can make is stick to an optimal motor mode to avoid fixing muscle hypo- and hypertonicity, which occur during long-term preservation of static postures, overloading by physical work and sports, etc. An individual should periodically change body positions without staying for a long time in any of them, uniformly load all the muscles in the body with non-intensive workload to prevent them from overwork or prolonged inactivity.

The SD is not a disorder itself, but a universal mechanism of the LS deterioration, linking together existing disorders and generating new ones: LS pathologies caused by external factors become IIFs of systemic destruction, other pathologies originate from it as its manifestations or consequences. Thus, via the SD mechanism, acute disorders become chronic, and chronic LS pathologies become systemic. No segment of the LS (neither the spine, nor any joint) is more important than the others (to think so is a common mistake), all LS segments equally participate in the general systemic destructive process. When considering the LS, one needs to remember that the LS always works as an integral system and always suffers also as an integral system. Therefore it is impossible to cure separately disorders of its parts without systemic influence on the whole LS. Moreover, as the LS systemic pathology is essentially organic and accumulates in the whole body during the whole life, it is impossible to eliminate it quickly or even instantly (by manipulations or surgery). In any case, it takes a long time (weeks, months) to restore the anatomically normal composition and properties of tissues of the elements in the whole LS.

Our results are not just a plain theory, they have direct practical implications. It is impossible to stop the SD, however the same mechanisms of adaptation and self-organization can be employed to counteract it. To solve this problem, on the basis of the LS systemic destruction theory we developed the world's first technology for diagnostics and correction of the LS anatomic state – the Systemic Reconstructive Therapy (SRT) [37,38]. The SRT intensifies the body self-repair processes and solves a broad range of problems of external correction of the LS anatomic state, which the body cannot solve on its own. During SRT development we discovered reversibility of organic changes in the LS elements tissues under external corrective influences.

The SRT does not treat specific disorders directly, but it is a universal technology for restoration and maintenance of the normal LS anatomic state. A rehabilitologist always acts systematically: regardless of the individual's specific complaints he/she always examines the whole LS (since the systemic organic pathology is present in every individual's LS) and then performs its systemic reconstruction trying to bring the LS state to the normal LS anatomic state as close as possible. The rehabilitologist creates the necessary conditions and controls the body self-repair process aimed at reversing organic changes in the LS elements tissues and their shape restoration, and gradually repositions LS elements in the spatial structure accordingly. As the LS state approaches the normal anatomic state, SD manifestations and consequences disappear automatically as the result of removal of the cause. Consequently, the functioning of LS elements and segments is normalized, a full range of their motion is restored, musculoskeletal pain disappears, LS deformities and visual body imperfections are eliminated, functioning of other organs and body systems is improved – the individual becomes much healthier in general. Such correction is possible at any age. The SD organically alters the whole human body, the SRT allows to alter it in the opposite direction using only human hands, without drugs, surgery or any device. Our experience has proven that the SRT is a highly effective physical rehabilitation technology for the human LS.

Perhaps not less, but even more important is that SRT may and should be applied, when the LS is still practically healthy and an individual has no complaints yet. This provides effective prevention of future disorders, elimination of their origins. Because the SD progresses in every individual's LS from the first years of life, it should be counteracted before severe manifestations appear. If the LS anatomic state is being corrected since the early childhood, it is possible to raise a healthy individual with a proportional and beautiful body, and then maintain it in this state, preventing emergence of pathologies.

The SD theory links the LS systemic pathology with pathologies of the nervous and cardiovascular systems, it is evident from Fig. 1 that they reinforce each other. Understanding of this interconnection may become the key to prevention of the cardiovascular system disorders that are the world's leading cause of death [39]. The LS is the largest system of the body amounting to the majority of its mass; it is a "house", where all the other organs and systems live. Muscles at rest contain 1/4 of the whole blood volume, 1/5 of the total blood flow passes through them, which dramatically increases during exercise and may exceed 4/5 of the total blood flow. That is why disturbance of blood circulation in muscles due to the LS systemic destruction affects blood and lymph circulation in the whole body. Blood stagnation takes place not only in muscles, but in the other LS elements, organs and body systems as well, pathologic deposits and thrombi can be formed in the vessels, inflammatory processes begin. This results in blood pressure disorders, headaches, somatoform autonomic dysfunction, chronic fatigue, constant malaise, and individual's irritability for no apparent cause; for this reason antidepressants are often prescribed to people who do not actually need them. Restoring normal circulation in the LS using the SRT, makes it possible to significantly improve the overall condition of the cardiovascular system and prevent occurrence of these and other more dangerous conditions such as heart attacks and strokes that is confirmed by our long-term practice.

The approach to study of the LS pathology based on an interdisciplinary synthesis of knowledge about functioning of complex systems that we applied, by analogy with the systems biology [40] and systems medicine [41], may be called systems orthopedics. The same approach can be used to study other subsystems of the body, which have similar principles of functioning. In these subsystems, adaptation and self-organization processes under the control of the corresponding control systems also take place in response to various disturbances. Early detection of deviations from the normal state of these subsystems and application of adequate external corrective actions (taking into account the control objectives of these subsystems) will enable elimination of causes of future pathologies, just like the SRT does. This new area of systems research of the locomotor system and the whole human body is now one of the most urgent and requires involvement of experts from various fields of science.